\documentclass[aps,prl,onecolumn,preprint,notitlepage,linenumbers]{revtex4-1}

\usepackage{graphicx} % support the \includegraphics command and options
\usepackage{epsfig}
\usepackage{epstopdf}
\usepackage{amsmath}
\usepackage{amsfonts}
\usepackage{amssymb} 
\usepackage{color}
\usepackage{multirow}
\usepackage{lineno}
\usepackage{url}
\begin{document}
\nolinenumbers
%\vspace*{-3cm}\hspace{4.5cm} IRFU-??-?? 
\title{Deeply virtual Compton scattering off the neutron} %\\: Flavor and Phase Analysis of Compton Form Factors}
\author{The Jefferson Lab Hall A Collaboration}
\affiliation{\vspace*{2cm}}

\date{\today}

%\modulolinenumbers[1]
\begin{abstract}
%\linenumbers
%The best way to probe the three dimensional structure of nucleons (protons and neutrons) is through the so-called Deeply Virtual Compton Scattering (DVCS), where a high energy electron is scattered off a nucleon which in turns emits a high energy photon. This photon is produced by one of the quarks inside the nucleon and carries information about its transverse position and longitudinal momentum. However, the photon, having no flavor, cannot tell about the nature (up or down) of the quark that emitted it. Here we report the first observation of DVCS off a neutron. Thanks to the approximate isospin symmetry of Quantum Chromodynamics, combining our measurements with previous ones off the proton, we can individually pin down different quark flavors inside the nucleon. %In particular, our findings indicate a significantly different 3D distribution of unpolarized up and down quarks inside a transversely polarized nucleon. This work opens a new avenue in the imaging of subatomic particles by allowing to probe their flavor content.
The three-dimensional structure of nucleons (protons and neutrons) is embedded in so-called generalized parton distributions, which are accessible from deeply virtual Compton scattering. In this process, a high energy electron is scattered off a nucleon by exchanging a virtual photon. Then, a highly-energetic real photon is emitted from one of the quarks inside the nucleon, which carries information on the quark's transverse position and longitudinal momentum. By measuring the cross-section of deeply virtual Compton scattering, Compton form factors related to the generalized parton distributions can be extracted. Here, we report the observation of unpolarized deeply virtual Compton scattering off a deuterium target. From the measured photon-electroproduction cross-sections, we have extracted the cross-section of a quasi-free neutron and a coherent deuteron. Due to the approximate isospin symmetry of quantum chromodynamics, we can determine the contributions from the different quark flavours to the helicity-conserved Compton form factors by combining our measurements with previous ones probing the proton's internal structure. These results advance our understanding of the description of the nucleon structure, which is important to solve the proton spin puzzle.
\end{abstract}

% insert suggested keywords - APS authors don't need to do this
%\keywords{}

%\maketitle must follow title, authors, abstract, \pacs, and \keywords
\maketitle

%%%%%%%%%%%%%%%%%
%\begin{linenumbers}
Understanding the quark and gluon structure of the nucleon is one of the outstanding challenges in hadronic physics. For decades, elastic and deep inelastic (DIS) lepton-nucleon scattering have been investigated to get information on the transverse spatial distribution of partons~\cite{Hofstadter:1955}, encoded by the nucleon form factors (FFs), and their longitudinal momentum distribution characterized by parton distribution functions (PDFs)~\cite{Taylor:1991}. The longitudinal direction is given by the momentum of the virtual photon mediating the interaction. By unifying FFs and PDFs, the theoretical framework of generalized parton distributions (GPDs)~\cite{Radyushkin:1997ki,Mueller:1998fv} allows a multi-dimensional description of the nucleon. In particular, GPDs correlate the transverse spatial structure of partons and their intrinsic longitudinal motion~\cite{Burkardt:2003},  offering the possibility to access the quark (and gluon) orbital angular momentum~\cite{Ji:1996ek} and to elucidate the nucleon spin puzzle~\cite{RevModPhys.85.655}. The nucleon structure is described by eight GPDs for each quark flavor $q$ at leading twist. The leading twist describes the scattering off a single parton of the nucleon with no other partons participating in the process. Four of the leading-twist GPDs conserve the helicity of the parton (chiral-even GPDs), denoted by $H^q$, $E^q$, $\widetilde{H}^q$ and $\widetilde{E}^q$, and the other four flip the parton helicity (chiral-odd GPDs), $H_T^q$, $E_T^q$, $\widetilde{H}_T^q$ and $\widetilde{E}_T^q$~\cite{Hoodbhoy:1998vm,Diehl:2003}. Each GPD depends on three variables, $x$, $\xi$ and $t$,  where $x+\xi$ ($x-\xi$) is the longitudinal momentum fraction of the struck quark before (after) the scattering resulting on a squared four-momentum transfer $t$ to the nucleon.

\begin{figure}
%    \centering
\begin{minipage}{0.65\linewidth}
    \includegraphics[width=0.9\linewidth]{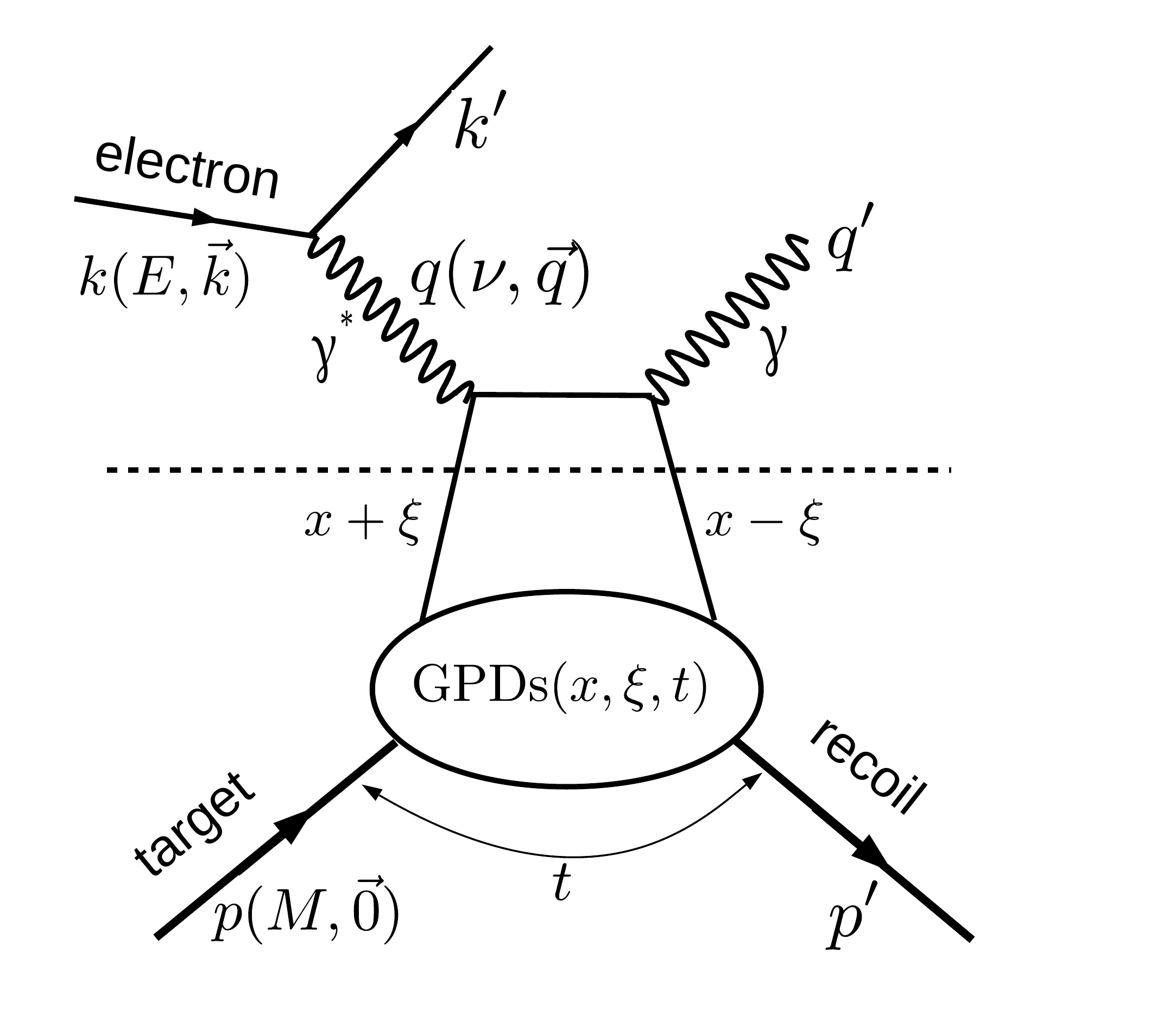}
    \end{minipage}\hfill\begin{minipage}{0.34\linewidth}
    \centerline{Invariants}
    \vskip -1.5em
    \begin{align*}
    Q^2 &= -(k-k')^2 \\
    x_B &= Q^2/(2q\cdot p) \\
%    W^2 &= (q+p)^2 \\
    y    &= (q\cdot p)/(k\cdot p)\\
    t      &= (q-q')^2  \\
    t^{\prime}      &= t_\text{min}-t  \\
    \end{align*}
    \end{minipage}
    \caption{The handbag diagram for DVCS on the nucleon ($M=M_N$) or the coherent deuteron ($M=M_d$). In the kinematics of the experiment reported here $x_B=0.36$ for DVCS on the nucleon and $x_B=0.18$ for DVCS on the coherent deuteron. 
    %The definitions of some kinematic variables are indicated in the right. 
    The minimal $|t|$ value is $t_{min} = Q^2 [2(1-x_B)(1-\sqrt{1+\epsilon^2})+\epsilon^2]/(4x_B(1-x_B)+\epsilon^2)$, where $\epsilon^2=4x_B^2M^2/Q^2$.}
    \label{figure1} %t_\text{min} formule old BKM (31) TBelitsky
\end{figure}

GPDs are involved in many hard exclusive processes such as Deeply Virtual Compton Scattering (DVCS) and Deeply Virtual Meson Production (DVMP), where a real photon and a meson are emitted respectively by the nucleon when probed with a virtual photon. In the Bjorken limit, when the virtuality $Q^2 $ and the energy $\nu$ of the virtual photon become very large at fixed $x_B$ (see Fig.~\ref{figure1}), QCD factorization theorems~\cite{Collins:1996fb,Ji:1998xh} demonstrate that the DVCS amplitude can be factorized into a hard perturbative kernel and a soft part described by chiral-even GPDs, leading to the so-called handbag diagram of Fig.~\ref{figure1}. Recent experimental studies on DVCS show that the Bjorken limit may already be reached at $Q^2$ values as low as 1.5 GeV$^2$~\cite{Defurne:2015kxq,Jo:2015ema}. Experimentally, DVCS is indistinguishable from Bethe-Heitler (BH) where the real photon is emitted by the incoming or the scattered electron.
%as shown in Fig.~\ref{figure2}
The differential cross section of photon electroproduction can then be written as~\cite{Belitsky:2010jw}:
%% \begin{figure}
%% \centering 
%% %\vspace*{5cm}
%% \includegraphics[width=0.7\linewidth]{fig2.pdf}
%% \caption{Lowest order QED diagrams for the process $eN \to eN \gamma$, including the DVCS and the Bethe-Heitler amplitudes.}
%% \label{figure2}
%% \end{figure}  
\begin{equation}
\frac{d^4\sigma}{dQ^2 dx_B dt d\phi}=\frac{\alpha_{QED}^3 ~x_B ~y^2}{8 \pi~ Q^4~ e^6~\sqrt{1+\epsilon^2 }}~
\Big[|\mathcal{T}_{BH}|^2+|\mathcal{T}_{DVCS}|^2 + \mathcal{I}
\Big]\;,
\label{sigtot}
\end{equation}
where $\phi$ is the azimuthal angle between the leptonic and hadronic planes~\cite{Bacchetta:2004jz}. The BH amplitude $\mathcal{T}_{BH}$ is fully calculable in QED with the nucleon FFs with 1\% uncertainty. The interference and $|\mathcal{T}_{DVCS}|^2$ terms in Eq.~(\ref{sigtot}) contain a finite $\cos(n\phi)$ series for $\mathcal{I}$ ($n=0\cdots 3$) and for $|\mathcal{T}_{DVCS}|^2$ ($n=0\cdots 2$)~\cite{Diehl:1997bu}. The different $\phi$ and beam-energy dependence of $\mathcal{I}$ and $|\mathcal{T}_{DVCS}|^2$ at fixed $x_B$, $Q^2$ and $t$ allows to deduce the individual contribution of these terms~\cite{Defurne:2017paw}. Their Fourier harmonics can be expressed respectively as a function of linear and bilinear combinations of GPD convolutions with the perturbative kernel, called Compton Form Factors (CFFs)~\cite{Belitsky:2014}. Photon electroproduction measurements at sufficiently high $Q^2$ are sensitive to different CFF combinations depending on the lepton and target polarization states~
\cite{Airapetian:2001yk,
  Stepanyan:2001sm,
  Adloff:2001cn,
  Chekanov:2003ya,
  Aktas:2005ty,
  Chen:2006na,
  Airapetian:2006zr,
  Mazouz:2007aa,
  Aaron:2007ab,
  Girod:2007aa,
  Airapetian:2008aa,
  Chekanov:2008vy,
  Gavalian:2008aa,
  Aaron:2009ac,
  Airapetian:2009aa,
  Airapetian:2010,
  Airapetian:2011uq,
  Airapetian:2012mq,
  Airapetian:2012pg,
  Jo:2015ema,
  Defurne:2015kxq,
  Defurne:2017paw,
  HirlingerSaylor:2018bnu}. While the incident beam helicity-dependent cross section can access the imaginary part of the interference, which is sensitive to GPDs at $x=\pm\xi$, the helicity-independent cross-section measurements offer a stronger constraint since the real part of the interference probes GPD integrals over their full $x-$domain. For instance, the combination of leading-twist CFFs $\mathcal{F}\in\{\mathcal{H},\mathcal{E},\mathcal{\widetilde{H}}\}$ appearing in the interference term for unpolarized target reads~\cite{Belitsky:2014,Braun:2014}
\begin{equation}
\Re e(C^{\mathcal{I}})= \Re e\bigg(F_1 \mathcal{H}-\frac{t}{4M^2}F_2\mathcal{E}+\xi(F_1+F_2)\mathcal{\widetilde{H}}\bigg)~,
\label{Ci}
\end{equation}
where $F_1(t)$ ($F_2(t)$) is the Dirac (Pauli) FF, $\xi\approx\frac{x_B}{2-x_B}$ is the skewness variable in the Bjorken limit, and 
\begin{equation}
\Re e(\mathcal{F})= \mathcal{P} \int_{-1}^{1} dx~\left[\frac{1}{x-\xi}\pm\frac{1}{x+\xi}\right]F(x,\xi,t)~.
\label{eq:ReCFF}
\end{equation}

Neutron GPDs are highly complementary to the proton ones. Their knowledge represents a mandatory step towards a better description of the partonic structure of the nucleon, even if their experimental measurement is more challenging to achieve. On the one hand, the neutron appears to be the simplest way to perform a flavor decomposition of the $u$ and $d$ quark GPDs. Indeed, the quark flavor structure of a given GPD $F$, neglecting strange quarks, writes for a proton (p) and a neutron (n)
\begin{eqnarray}
F^{p,n}=\frac{4}{9}F^{u,d}+\frac{1}{9}F^{d,u}~.
\label{flavor}
\end{eqnarray}
Contributions from strange quark GPDs are expected to be negligible based on the size of strange flavor parton distribution functions and the strange vector and axial-vector strange form factors of the nucleon, all of them very small compared to their $u$ and $d$ counterparts~\cite{Dulat:2015mca,Maas:2017snj}. On the other hand, the different FF values for a neutron and a proton allow a sensitivity to specific CFFs. For instance, Eq.~(\ref{Ci}), which is dominated by $\mathcal{H}$ and $\mathcal{\widetilde{H}}$ for the proton, becomes mainly sensitive to $\mathcal{E}$ for the neutron due to the small value of $F_1$ and to the cancellation between $u$ and $d$ polarized parton distributions in $\mathcal{\widetilde{H}}$~\cite{Goeke:2001}. Measurements of the unpolarized $en\to en\gamma$ (n-DVCS) cross sections at low $t$ become then of direct relevance in the determination of the quark angular momentum via Ji's sum rule~\cite{Ji:1996ek}:
\begin{equation}  
  J^{q}=\frac{1}{2} \int_{-1}^1 xdx [ H^{q}(x,\xi,t=0)+E^{q}(x,\xi,t=0)] ~~ \forall \xi,
\label{Ji} 
\end{equation}
since the $x$ dependence of $E^q$ is basically unknown, contrary to $H^q$. The only existing n-DVCS data at large $x_B$ are from the pioneer JLab-Hall A experiment E03-106~\cite{Mazouz:2007aa},  where the beam helicity-dependent cross section was determined at $x_B=0.36$, $Q^2=1.9$ GeV$^2$ and $E=5.75$ GeV. In this paper, we present the first $en\to en\gamma$ unpolarized cross-section measurements at a very close kinematics ($x_B=0.36$ and $Q^2=1.75$ GeV$^2$) for two beam energies $E=4.45$ and 5.55 GeV. 

The data of the E08-025 experiment reported herein were acquired in JLab--Hall A. The extraction of the $n(e,e\gamma)n$ cross section in the quasi-free approximation is based on a controlled subtraction of data taken on liquid hydrogen (LH2) and liquid deuterium (LD2) targets, similarly to what was done in Ref.~\cite{Mazouz:2007aa} and more recently in Ref.~\cite{Mazouz:2017skh}. The quasi-free $p(e,e\gamma)p$ contribution is determined from the data of the E07-007 experiment~\cite{Defurne:2017paw}, the LH2 and LD2 targets being daily switched to minimize systematic errors. Scattered electrons were detected in the left High Resolution Spectrometer (HRS)~\cite{Alcorn:2004sb} of Hall A, defining accurately the leptonic variables and the interaction vertex. Photons of more than 500 MeV were detected in an electromagnetic calorimeter consisting of a $13\times 16$ array of $3\times3\times18.6$ cm$^3$ PbF$_2$ crystals, placed at 1.1 m from the target and centered around the virtual photon direction. This configuration leads to a $2\pi$ coverage in $\phi$ and a momentum transfer $t$ ranging from -0.15 to -0.45 GeV$^2$ ($t^{\prime}=t_{min}-t$ $\in [0, 0.3]$ GeV$^2$). 

Figure~\ref{figure2} (top) shows the missing mass squared $M_X^2=(q+p-q')^2$ spectrum  of D$(e,e'\gamma)X$ where the target is assumed to be a nucleon at rest ($p=(M_N,\vec{0})$). Exclusive photon electroproduction events are located around $M_N^2$ and are contaminated by three kinds of events: accidentals, photons coming from $\pi^0$ decays and semi-inclusive (SIDIS),
equivalently associated DVCS~\cite{Guichon:2003ah}, events $eN\to e\gamma X$. 
The accidentals are determined by analyzing events that are outside the coincidence window. Their relative contribution is reduced by applying the selection criterion $M_X^2>0.5$~GeV$^2$ on the data set. The number of $\pi^0$ decays yielding only one photon in the calorimeter acceptance is estimated by generating thousands of decays for each detected $\pi^0$~\cite{Defurne:2015kxq}. The subtracted number of events $N^i_{\pi^0}$ in a bin $i$ is :
\begin{equation}
  N^i_{\pi^0}={\sum_{j=1}^{N_{\pi^0}}}\displaystyle\, \frac{N^{j,i}_1}{N^{j}_2} \,\pm\, \textstyle\sqrt{\sum_{j=1}^{N_{\pi^0}}(\frac{N_1^{j,i}}{N^{j}_2})^2}\,,
  \label{eq:pi0}
\end{equation}
where the sum runs over the total number $N_{\pi^0}$ of detected $\pi^0$ events. For each of those $j$ events, $N^{j,i}_1$ is the number of simulated decays yielding one photon in bin $i$ and fulfilling the DVCS selection criteria, whereas $N^j_2$ is the number of decays that yield two photons within the experimental acceptance. 
%experimental acceptance when one of the decays (from the same detected $\pi^0$ event) yields only one photon in the experimental acceptance, in bin $i$. The first factor in Eq.~\ref{eq:pi0} accounts for the number $N_{dec}$ of decays generated for each detected $\pi^0$ event, while the second factor under the summation sign makes up for the fact that only a fraction $N^i_2/N_{dec}$ of events are used in the calculation. %Selection criteria are applied to ensure the same acceptance for DVCS photons and $\pi^0$ decays so that the normalization of the background subtraction is only dependent on the number of generated decays for each of the detected $\pi^0$ events.
%%%
%%%
The SIDIS events are essentially located above the pion production threshold $(M_N+M_{\pi})^2\approx 1.15$~GeV$^2$. However, the nominal selection criterion $M_X^2<0.95$~GeV$^2$ is applied to minimize any possible contamination to the exclusive yield due to resolution effects. A bin-dependent systematic error is attributed later to the results by studying the stability of the extracted cross sections when varying this nominal selection criterion. After the subtraction of accidentals and the $\pi^0$ background, the remaining events in the exclusive region can be decomposed, within the impulse approximation, into coherent elastic events $d(e,e^{\prime}\gamma)d$ and two incoherent quasi-elastic channels
\begin{equation}
D(e,e^{\prime}\gamma)X = d(e,e^{\prime}\gamma)d + n(e,e^{\prime}\gamma)n + p(e,e^{\prime}\gamma)p , 
\label{impulse}
\end{equation}
where $X = np \oplus d$. The extraction of the $ed\to ed\gamma$ cross section is also considered in this work. Its expression is similar to Eq.~(\ref{sigtot}) and depends on deuteron CFFs involving, at leading twist, nine spin-1 target GPDs~\cite{Kirchner2003}. The quasi-free $p(e,e^{\prime}\gamma)X$ contribution is determined by normalizing the LH2 data to the luminosity of the LD2 data and by adding statistically the Fermi-momentum~\cite{Lacombe:1980} of bound protons inside the deuteron. The width variation of the $M_X^2$ distribution due to the Fermi-momentum smearing is less than 1\% and thus its uncertainty negligible on the final results. Figure~\ref{figure2} (bottom) shows the result of the subtraction of the background-free LH2 from the LD2 data. The resulting events passing the $M_X^2$ exclusivity selection criterion correspond to the $d(e,e^{\prime}\gamma)d$ and $n(e,e^{\prime}\gamma)n$ channels, which are kinematically separated by $\Delta M_X^{2} =t(1-M_N/M_d)\approx t/2$. The exclusive data are divided into $12\times 2 \times 5 \times 30$ bins in $\phi$, $E$, $t$ and $M_X^{2}$ respectively. The results will be presented as a function of $\phi$, $E$ and $t$, the binning on $M_X^{2}$ serving only to separate the $d(e,e^{\prime}\gamma)d$ and $n(e,e^{\prime}\gamma)n$ contributions by exploiting their kinematic shift.

%The cross sections are extracted by fitting the experimental number of counts to the simulated number of events in each experimental bin. The simulated yield is parametrized as a function of different CFFs that whose values are fitted in order to minimize a $\chi^2=\sum_i(N^{data}-N^{sim})^2/\sigma_{data}^2$, where the sum runs over all experimental bins $i$.

The extraction of the cross sections is based on a simultaneous fit of all experimental bins by means of a Monte-Carlo simulation of $d(e,e^{\prime}\gamma)d$ and $n(e,e^{\prime}\gamma)n$ reactions. %in which the cross sections are parametrized by the different CFFs. 
After applying the exclusivity $M_X^2$ selection criterion, the remaining number of simulated events $N^{\text{sim}}_i$ in bin $i$ is adjusted to the corresponding number of experimental events $N^{\text{exp}}_i$ by minimizing% the $\chi^2$:
\begin{equation}
\chi^2=\sum_{i=1}^{3600}\left(\frac{N^{exp}_i-N^{sim}_i}{\delta^{exp}_i}\right)^2\;,
\label{eq::fit}
\end{equation}
where $\delta^{\text{exp}}_i$ is the statistical uncertainty of $N^{\text{exp}}_i$. The free parameters of the fit are a set of neutron and deuteron CFF combinations for each $t$ bin. %The variation of the results when adding possible twist-3 CFF combinations in the minimization procedure is treated as an additional source of systematic error.
Figure~\ref{figure3} presents the $\phi$-dependent cross sections for both beam energies and for all $t$ bins excepting the highest $|t|$ one, which was only used to account for bin migration effects. The uncertainties on the extracted neutron and coherent deuteron cross sections take into account the $\Delta M_X^{2}$ correlation between these two contributions, which varies between -0.96 at low $|t|$ and -0.79 at high $|t|$. Point-to-point systematic uncertainties due to calorimeter calibration, the simulation smearing and the missing mass selection criterion are added in quadrature to a 3.1\% normalisation uncertainty (see Methods) and are represented by the boxes around the data points in Fig.~\ref{figure3}. Note that the normalization uncertainties (3.1\%) are much smaller than the point-to-point uncertainties (which average to 37\%) and their effect is negligible.

%\begin{figure}
%\centering
%\includegraphics[width=\linewidth]{bin2_nbreEvt.pdf}
%\caption{Total events number as a function of $\phi$ (with statistical uncertainty) at $E =4.45$ GeV (left) and $E=5.55$ GeV (right), in the bin $\left<t^{\prime}\right>=0.163\text{ GeV}^2$ (neutron kinematics), equivalently $\left<t^{\prime}\right>$=0.134 GeV$^2$ (deuteron kinematics).  The blue (red) histogram shows the neutron (deuteron) Bethe-Heitler contribution and the solid black histogram represents the sum of both histograms (Bethe-Heitler of neutron +  Bethe-Heitler of deuteron).}% at   $x_B=0.36$ ($x_d=0.18$). }
%\label{figure3}
%\end{figure}

When integrated over $\phi$, the neutron results exhibit a significant deviation from the BH contribution, especially for $E=4.45$~GeV, as shown in Fig.~\ref{figure4}. At $E=4.45$~GeV the coherent deuteron cross sections are smaller than the neutron cross sections for the two highest $|t|$ bins, as could be expected from Fig.~\ref{figure2} and by the rapid relative decrease of the deuteron FFs at large $|t|$. The coherent deuteron results are relatively well described, within uncertainties, by theoretical calculations based on deuteron GPDs~\cite{canopire}, whereas the VGG model~\cite{Vanderhaeghen:1998uc, VGG} overshoots significantly results for the neutron. This model reproduces proton DVCS data much better (see~\cite{Jo:2015ema} for example), so the disagreement for the neutron 
%may indicate that GPD $E$, which can hardly be constrained by proton data, is challenging to model.
is perhaps symptomatic of the paucity of experimental constraints on the GPD $E$.

%especially at high $|t|$, where the measurement uncertainties are relatively small. The coherent deuteron cross sections are found to be compatible with the corresponding BH for the lowest $|t|$ bin within their large uncertainties. The coherent deuteron results are compatible, within uncertainties, with theoretical calculations based on deuteron GPDs~\cite{canopire} for all bins.

%%%%%%%%%%%
%%%%%%%%%%%%%%%%%

The simultaneous fit of both beam energy settings allows to independently extract the contribution from the $|\mathcal{T}_{DVCS}|^2$ and BH-DVCS interference terms. These are shown in Fig.~\ref{figure5} for the neutron. The analysis has been performed within the recent formalism by Braun et al.~\cite{Braun:2014} which accounts for kinematical power corrections of $\mathcal O(t/Q^2)$ and $\mathcal O(M^2/Q^2)$. Previous results on the proton~\cite{Defurne:2017paw} showed the necessity to include higher twist (HT) or next-to-leading order (NLO) CFFs in the analysis in order to accurately reproduce the azimuthal angular dependence of the cross section. Fits have been performed within two scenarios that yield equally good results. A higher-twist scenario includes, in addition to the helicity-conserving CFFs $\mathcal H_{++}$, $\widetilde{\mathcal H}_{++}$ and $\mathcal E_{++}$, the HT CFFs $\mathcal H_{0+}$, $\widetilde{\mathcal H}_{0+}$ and $\mathcal E_{0+}$. The next-to-leading order scenario includes helicity-conserving CFFs and the NLO CFFs $\mathcal H_{-+}$, $\widetilde{\mathcal H}_{-+}$ and $\mathcal E_{-+}$~\cite{Braun:2014}. The absence of $\mathcal{\widetilde{E}}$ in the interference term makes it difficult to separate its real and imaginary parts and was not included in the fit. The separation of the $|\mathcal{T}_{DVCS}|^2$ and interference terms in Fig.~\ref{figure5} shows slight variations depending on which scenario is considered, but with a significant signal from $|\mathcal{T}_{DVCS}|^2$ for all values of $\phi$ and $t$.

The notable size of the $|\mathcal{T}_{DVCS}|^2$ and interference terms constitutes the first observation of DVCS off a quasi-free neutron target and allows, when combined with results off the proton, to probe nucleon GPDs at the level of quark flavors $u$ and $d$ by exploiting Eq.~(\ref{flavor}). A fit of all the available proton~\cite{Defurne:2015kxq,Defurne:2017paw} and neutron~\cite{Mazouz:2007aa} cross sections from Hall A at $Q^2=1.5-2.3$ GeV$^2$ and $x_B=0.36$, including the present results, is performed within the BMMP parametrization~\cite{Braun:2014} of the DVCS amplitude. Fits were performed in the same two scenarios described above, HT and NLO, but for each flavor of quarks $u$ and $d$. The fits quality is quite reasonable across the whole data set (see Extended Data Figures 1 and 2) with $\chi^2/ndf$ ranging from $399/444$ ($407/444$) to $533/470$ ($529/470$) for the HT (NLO) scenario.
Figure~\ref{figure6} shows the results for the real and imaginary parts of $\mathcal{H}_{++}$, $\mathcal{\widetilde{H}}_{++}$ and $\mathcal{E}_{++}$. The better accuracy of the proton experimental cross sections is reflected in the $u$ CFF results. Notice that the uncertainties in Fig.~\ref{figure6} are dominated by correlations in the fit parameters rather than by the accuracy of the experimental cross-section values. 

The experimental results in Fig.~\ref{figure6} are compared to theoretical predictions based on a reggeized diquark model of GPDs~\cite{Goldstein:2010gu, Goldstein:2013gra}. The flavors $u$ and $d$ for CFF $\mathcal H_{++}$ show the same sign, for both the real and imaginary parts, in the model. Data are also consistent, within uncertainties, with the same sign of $u$ and $d$ for $\mathcal H_{++}$. This is in agreement with what is observed in the forward limit~\cite{Dulat:2015mca} and the predictions of SU($N_c$) gauge theory with large number $N_c$ of colors~\cite{Goeke:2001}. In both the forward and the large-$N_c$ limits GPD $\widetilde{H}$ has opposite signs for flavors $u$ and $d$ which seems to be also the case for the data. Notice that there is a change of sign between the imaginary and the real parts of $\widetilde{\mathcal{H}}$. While in this case this effect is well reproduced by the model, it highlights the non-trivial functional form of the GPDs, which can flip the sign of the integral defining the real part of the CFFs (Eq.~(\ref{eq:ReCFF})).
The CFF $\mathcal{E}_{++}$ shows interesting features. Its imaginary parts suggest opposite signs for flavors $u$ and $d$. This matches the signs of the $u$ and $d$ anomalous magnetic moments $\kappa^{u,d}$, which give the normalization of the first moment of $E$ and also agrees with the large $N_c$ limit~\cite{Goeke:2001}. However, the values for quark $u$ are not well reproduced by the theoretical model. This is also the case, to some extent, for $\Im m(\mathcal H_{++})$, $\Re e(\mathcal {\widetilde{H}_{++}})$ and $\Im m(\mathcal {\widetilde{H}_{++}})$. %The real part of $\mathcal{E}_{++}$ show uggests the same sign for $u$ and $d$, indicating a non-trivial flavor difference to the $x-$dependences of the GPDs $E_{u,d}$, which in this case is not reproduced by the theory.
%  CEH insert
The Ji sum rule (Eq.~\ref{Ji}) results of the model in Ref.~\cite{Goldstein:2010gu} are presented in \cite{GonzalezHernandez:2012jv}
(Fig.~19 and Table V).
This model is flexible, and since there are few constraints on CFF $\mathcal{E}_{++}$,  it is likely that a revision of
the model parameters would improve the description of these data, and result in a shift in the estimated values
of $J_{u,d}$ -- the contribution of up- and down-quarks to the spin of the proton.
%
%adjust the up-quark contribution, without significantly effecting the agreement with other observables.  This would, in turn, result
%in a larger estimated value for $J_u$ --- the contribution of up-quarks to the spin of the proton.

The forward limit of GPD $E$ is not measurable from any known inclusive process, and very few observables (such as DVCS off the neutron, and DVCS on a transversely polarized target) have high sensitivity to it. While the uncertainties in this first flavor separation presented here are still large, the data clearly show the potential and the sensitivity to this very challenging yet fundamental quantity. The upcoming  and in-process
experiments on proton and neutron DVCS with Jefferson Lab at 12 GeV will soon allow to better pin down the GPD $E$ and its flavor decomposition.

\section{Acknowledgements} We acknowledge essential work of the JLab accelerator staff and the Hall A technical staff. This work was supported by the Department of Energy (DOE), the National Science Foundation, the French {\em Centre National de la Recherche Scientifique}, the {\em Agence Nationale de la Recherche}, the {\em Commissariat \`a l'\'energie atomique et aux \'energies alternatives} and P2IO Laboratory of Excellence. Jefferson Science Associates, LLC, operates Jefferson Lab for the U.S. DOE under U.S. DOE contract DE-AC05-060R23177.

%%%%%%%%%%%
\section{Author information}
C.W. de Jager, S. Frullani, A. Saha and P. Solvignon are deceased
\subsection{Affiliations}
\begin{itemize}
\item Facult\'e des Sciences de Monastir, Avenue de l'environnement, 5019, Monastir, Tunisia\\
M. Benali, M. Mazouz

\item Institut de Physique Nucl\'eaire CNRS-IN2P3, 15 rue Georges Cl\'emenceau, 91406, Orsay, France\\
C. Desnault, C. Mu\~noz Camacho \& R. Paremuzyan, A. Mart\'i Jiménez-Arg\"uello

\item Syracuse University, 900 South Crouse Ave., Syracuse, NY, 13244, USA\\
Z. Ahmed \& A. Rakhman

\item Texas A\&M University-Kingsville, Engineering Complex, 700 University Blvd, Kingsville, TX, 78363, USA\\
H. Albataineh

\item Massachusetts Institute of Technology, 77 Massachusetts Ave, Cambridge, MA, 02139, USA\\
K. Allada, J. Huang \& V. Sulkosky

\item California State University, 5151 State University Dr, Los Angeles, CA, 90032, USA\\
K. A. Aniol, S. Iqbal \& D. J. Margaziotis

\item INFN/Sezione di Catania, Via S. Sofia, 62, 95125, Catania, Italy\\
V. Bellini, A. Giusa, G. Russo, M. L. Sperduto \& C. Sutera

\item Clermont Universit\'e, Universit\'e Blaise Pascal, CNRS/IN2P3, 4 Avenue Blaise Pascal, 63178, Aubire Cedex, France\\
M. Benali, P. Bertin, M. Brossard, E. Fuchey, C. E. Hyde, F. Itard, M. Magne \& C. Mu\~noz Camacho

\item Florida International University, 11200 SW 8th St, Miami, FL, 33199, USA\\
W. Boeglin \& P. Markowitz

\item Thomas Jefferson National Accelerator Facility, 12000 Jefferson Ave, Newport News, VA, 23606, USA\\
P. Bertin, A. Camsonne, J.-P. Chen, C. W. de Jager, A. Deur, R. Ent, D. Gaskell, J. Gomez, O. Hansen, D. Higinbotham, C. Keppel, J. J. LeRose, D. Meekins, R. Michaels, P. Nadel-Turonski, Y. Qiang, A. Saha, B. Sawatzky, P. Solvignon, B. Wojtsekhowski \& J. Zhang

\item Old Dominion University, Norfolk, 5115 Hampton Blvd, Norfolk, VA, 23529, USA\\
M. Canan, C. E. Hyde, S. Koirala \& M. N. H. Rashad

\item Ohio University, 123 University Terrace, 1 Ohio University, Athens, OH, 45701, USA\\
S. Chandavar \& J. Roche

\item Hampton University, 100 E Queen St, Hampton, VA, 23668, USA\\
C. Chen \& N. Nuruzzaman

\item Irfu, CEA, Universit\'e Paris-Saclay, 91191, Gif-sur-Yvette, France\\
M. Defurne, E. Fuchey \& F. Sabati\'e

\item Universit\`a di Bari, Piazza Umberto I, 1, 70121, Bari, Italy\\
R. de Leo

\item Rutgers, The State University of New Jersey, 7 College Ave, New Brunswick, NJ, 08901, USA\\
L. El Fassi

\item Temple University, 1801 N Broad St, Philadelphia, PA, 19122, USA\\
D. Flay, B. Sawatzky \& H. Yao

\item Carnegie Mellon University, 5000 Forbes Ave, Pittsburgh, PA, 15213, USA\\
M. Friend

\item University of Connecticut, 2390 Alumni Drive, Unit 3206, Storrs, CT, 06269, USA\\
E. Fuchey

\item INFN/Sezione Sanit\`a, Viale Regina Elena 299, 00161, Roma, Italy\\
S. Frullani, F. Garibaldi \& F. Meddi

\item Kharkov Institute of Physics and Technology, Akademichna St, 1, Kharkov, Kharkiv Oblast, 61000, Ukraine\\
O. Glamazdin

\item North Carolina Central University, 1801 Fayetteville St, Durham, NC, 27707, USA\\
S. Golge

\item Longwood University, 201 High St, Farmville, VA, 23909, USA\\
T. Holmstrom

\item The Catholic University of America, 620 Michigan Ave NE, Washington, DC, 20064, USA\\
T. Horn

\item Duke University, Physics Bldg., Science Dr., Campus Box 90305, Durham, NC, 27708, USA\\
M. Huang

\item Seoul National University, 1 Gwanak-ro, Gwanak-gu, Seol, South Korea\\
H. Kang

\item College of William and Mary, Department of Physics, P.O. Box 8795, Williamsburg, VA, 23187, USA\\
A. Kelleher \& B. Zhao

\item Tel Aviv University, P.O. Box 39040, Tel Aviv, 6997801, Israel\\
I. Korover

\item University of Virginia, 382 McCormick Rd, Charlottesville, VA, 22904, USA\\
R. Lindgren, K. Saenboonruang, W. A. Tobias, D. Wang, Z. Ye, Z. Zhao, X. Zheng \& P. Zhu

\item Kent State University, 800 E Summit St, Kent, OH, 44240, USA\\
E. Long \& L. Selvy

\item University of Massachusetts, 1126 Lederle Graduate Research Tower (LGRT), Amherst, MA, 01003, USA\\
J. Mammei

\item Facultad de F\'isica, Universidad de Valencia, Carrer del Dr. Moliner 50, 46100, Burjassot, Spain\\
A. Mart\'i Jiménez-Argüello

\item University of Ljubljana, Kongresni trg 12, 1000, Ljubljana, Slovenia\\
M. Mihovilovic \& S. Sirca

\item Los Alamos National Laboratory, Los Alamos, NM, 87545, USA\\
A. Puckett

\item Norfolk State University, 700 Park Avenue, Norfolk, VA, 23504, USA\\
V. Punjabi

\item Stony Brook University, 100 Nicolls Rd, Stony Brook, NY, 11794, USA\\
S. Riordan

\item Kasetsart University, 50 Thanon Ngam Wong Wan, Khwaeng Lat Yao, Khet Chatuchak, Krung Thep, Maha Nakhon, 10900, Thailand\\
K. Saenboonruang

\item Yerevan Physics Institute, 2. Alikhanian Br. Street, Yerevan, 0036, Armenia\\
A. Shahinyan

\item University of New Hampshire, 105 Main St, Durham, NH, 03824, USA\\
P. Solvignon

\item George Washington University, 2121 I St NW, Washington, DC, 20052, USA\\
R. Subedi

\item INFN/Sezione di Roma, Piazzale Aldo Moro 2, 00185, Roma, Italy\\
G. M. Urciuoli

\item Argonne National Laboratory, 9700 Cass Ave, Lemont, IL, 60439, USA\\
  X. Zhan
\end{itemize}

\subsection{The Jefferson Lab Hall A Collaboration}

M.~Benali,
C.~Desnault,
M.~Mazouz,
Z.~Ahmed,
H.~Albataineh,
K.~Allada,
K.~A.~Aniol,
V.~Bellini,
W.~Boeglin,
P.~Bertin,
M.~Brossard,
A.~Camsonne,
M.~Canan,
S.~Chandavar,
C.~Chen,
J.-P.~Chen,
M.~Defurne,
C.W.~de~Jager,
R.~de~Leo,
A.~Deur,
L.~El~Fassi,
R.~Ent,
D.~Flay,
M.~Friend,
E.~Fuchey,
S.~Frullani,
F.~Garibaldi,
D.~Gaskell,
A.~Giusa,
O.~Glamazdin,
S.~Golge,
J.~Gomez,
O.~Hansen,
D.~Higinbotham,
T.~Holmstrom,
T.~Horn,
J.~Huang,
M.~Huang,
G.M.~Huber,
C.E.~Hyde,
S.~Iqbal,
F.~Itard,
Ho.~Kang,
Hy.~Kang,
A.~Kelleher,
C.~Keppel,
S.~Koirala,
I.~Korover,
J.J.~LeRose,
R.~Lindgren,
E.~Long,
M.~Magne,
J.~Mammei,
D.J.~Margaziotis,
P.~Markowitz,
A.~Mart\'i Jim\'enez-Arg\"uello,
F.~Meddi,
D.~Meekins,
R.~Michaels,
M.~Mihovilovic,
N.~Muangma,
C.~Mu\~noz~Camacho,
P.~Nadel-Turonski,
N.~Nuruzzaman,
R.~Paremuzyan,
R.~Pomatsalyuk,
A.~Puckett,
V.~Punjabi,
Y.~Qiang,
A.~Rakhman,
M.N.H.~Rashad,
S.~Riordan,
J.~Roche,
G.~Russo,
F.~Sabati\'e,
K.~Saenboonruang,
A.~Saha,
B.~Sawatzky,
L.~Selvy,
A.~Shahinyan,
S.~Sirca,
P.~Solvignon,
M.L.~Sperduto,
R.~Subedi,
V.~Sulkosky,
C.~Sutera,
W.A.~Tobias,
G.M.~Urciuoli,
D.~Wang,
B.~Wojtsekhowski,
H.~Yao,
Z.~Ye,
L.~Zana,
X.~Zhan,
J.~Zhang,
B.~Zhao,
Z.~Zhao,
X.~Zheng,
P.~Zhu

%%%%%%%%%%%%%

\subsection{Author contributions.}
%M.B., C.D., M.M., Z.A., H.A., K.A., K.A.A., V.B., W.B., P.B., M.B., A.C., M.C., S.C., C.C., J.-P.C. M.D., C.W. de J., R. de L., A.D., L.E.F., R.E., D.F., M.F., E.F., S.F., F.G., D.G., A.G., O.G., S.G., J.G., O.H., D.H., T.H., T.H., J.H., M.H., G.M.H., C.E.H., S.I., F.I., Ho.K., Hy.K., A.K., C.K., S.K., I.K., J.J.L., R.L., E.L., M.M., J.M., D.J.M., P.M., A.M.J.-A., F.M., D.M., R.M., M.M., N.M., C.M.C., P.N.-T., N.N., R.P., R.P., A.P., V.P., Y.Q., A.R., M.N.H.R., S.R., J.R., G.R., F.S., K.S., A.S., B.S., L.S., A.S., S.S., P.S., M.L.S., R.S., V.S., C.S., W.A.T., G.M.U., D.W., B.W., H.Y., Z.Y., L.Z., X.Z., J.Z., B.Z., Z.Z., X.Z. and P.Z. 

The Jefferson Lab Hall A Collaboration constructed and operated the experimental equipment used in this experiment. Data were taken by a large number of collaboration members. The authors who performed data analyses and Monte Carlo simulations were M. Benali, C. Desnault, M. Mazouz and C.~Mu\~noz Camacho. The main authors of this manuscript were M. Benali, M. Mazouz, C. Mu\~noz Camacho, C. Hyde, J. Roche and A. Camsonne. It was reviewed by the entire collaboration before publication, and all authors approved the final version of the manuscript.

\section{Competing interests} The authors declare no competing interests.

\section{Data availability} Data that support the findings of this study are publicly available in \\
{\color{blue} https://userweb.jlab.org/$\sim$mazouz/NP/}

\section{Code availability} The computer codes that support the plots within this paper and the findings of this study are available from M.M. upon request.

%\appendix
\section{Methods}

A thorough bin-dependent monitoring of the experimental calibration and resolution between LH2 and LD2 data is performed to ensure a proper subtraction of the $p(e,e^{\prime}\gamma)p$ contribution from the exclusive $D(e,e^{\prime}\gamma)X$ yield. This monitoring is based on the reconstruction of the $\pi^0$ and the nucleon squared masses from respectively the 2-photon invariant mass $(q_1+q_2)^2$ (where $q_1$ and $q_2$ are the four-momenta of each photon) and the missing mass $(q+p-q_1-q_2)^2$ distributions in $\pi^0$ electroproduction events~\cite{mazouz_nst}. The energy calibration coefficients are adjusted to best reproduce the values of the $\pi^0$ and nucleon masses in each bin in $t$ and $\phi$. Similarly, the energy resolution of events in the LH2 data is adjusted for each of the bins in order to match the $\pi^0$ and nucleon mass resolutions observed in the LD2 data.

\begin{figure}
\centering 
\includegraphics[width=0.5\linewidth]{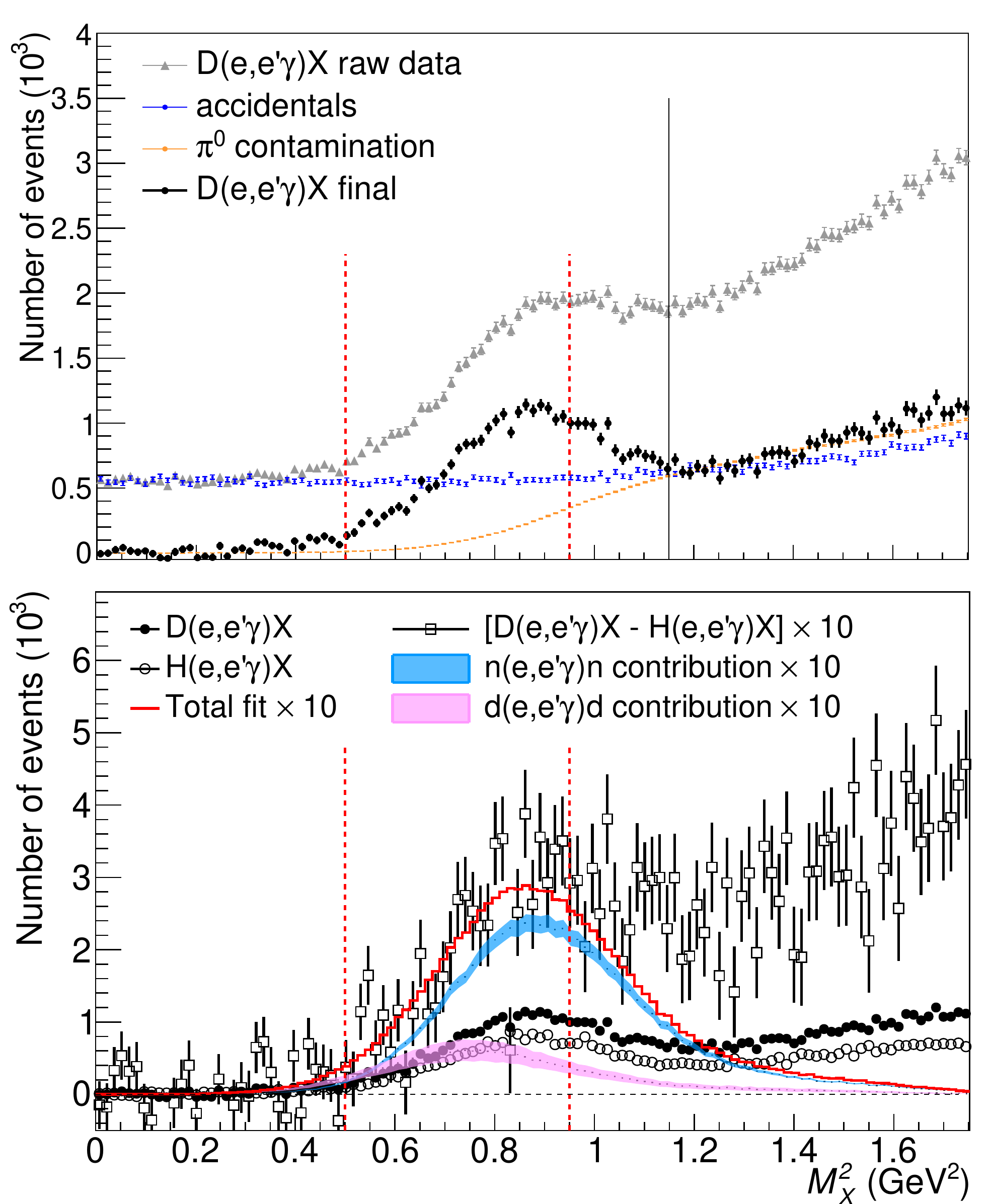}
\caption{Missing mass squared distributions. (Top) The grey triangles show the raw data distribution of D$(e,e'\gamma)X$ for $E=4.45$ GeV and the bin $\left<t\right>=-0.32$~GeV$^2$, integrated over $\phi$. The contributions of accidentals and $\pi^0$ contamination are shown in blue and orange respectively. The subtraction of these two contributions from the raw spectrum yields the black circles histogram (also shown in the bottom plot). The error bars of the raw data and the accidentals contribution correspond to standard deviations (s.d.) and are calculated as the squared root of the number of detected events. The error bars of the $\pi^0$ contamination contribution are calculated following Eq.~\ref{eq:pi0}. The pion production threshold is represented by the solid vertical line at 1.15~GeV$^2$. The range in $M_X^2\in[0.5,0.95]$~GeV$^2$ used in the analysis is shown by the dashed vertical lines. (Bottom) The difference between the D$(e,e'\gamma)X$ (black circles) and normalized Fermi-smeared H$(e,e'\gamma)X$ events (white circles), after accidental and $\pi^0$ background subtraction, is shown by the white squares histogram (scaled by a factor 10 for clarity). The blue and magenta bands (both scaled $\times 10$), show the simulated $n(e,e^{\prime}\gamma)n$  and $d(e,e^{\prime}\gamma)d$ yields, respectively, fit to the data by minimizing Eq.~(\ref{eq::fit}). These bands include the s.d. statistical uncertainty of the fit. The total fit to the white squares distribution is shown by the red histogram.}
\label{figure2}
\end{figure}

\begin{figure}
\centering
\includegraphics[width=0.6\linewidth]{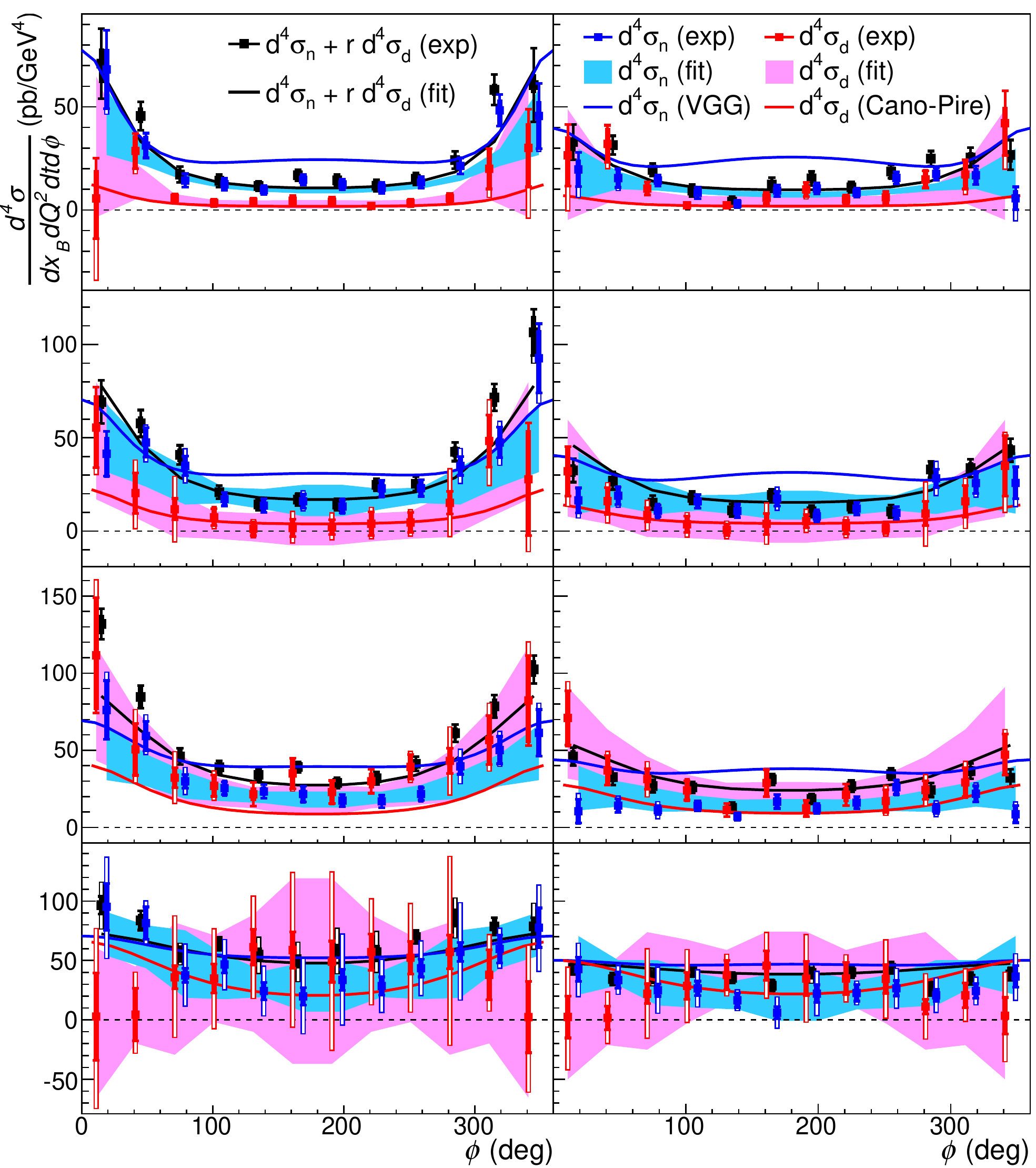}
\caption{Beam-helicity-independent cross sections. The black points show the sum of the neutron and coherent deuteron cross sections $d^4\sigma^n +r~d^4\sigma^d$ where $r=(dx_B^ddt^d)/(dx_B^ndt^n)\approx\frac{M_N}{M_d}\approx0.5$ is the ratio of the deuteron and neutron acceptances. The error-bars show the s.d. statistical uncertainty and the boxes around the points show the total s.d. systematic uncertainty. The blue (red) points show the neutron (coherent deuteron) contribution $d^4\sigma^n$ ($d^4\sigma^d$) with their s.d. statistical (bars) and systematics (boxes) uncertainties. The blue and magenta bands show the fit to  $d^4\sigma^n$ and $d^4\sigma^d$ respectively with the s.d. systematic and statistical errors of the fit added quadratically. The results correspond to $x_B=0.36$ for the neutron and $x_B=0.18$ for the coherent deuteron at $E =4.45$ GeV (left) and $E=5.55$ GeV (right). From top to bottom, the squared momentum transfer corresponds to $-\left<t\right>=$0.40, 0.32, 0.25 and 0.18 GeV$^2$ for the neutron and $-\left<t\right>=$0.33, 0.26, 0.20 and 0.15 GeV$^2$ for the deuteron. The solid blue (red) lines are theoretical calculations for the neutron (coherent deuteron) from Ref.~\cite{Vanderhaeghen:1998uc, VGG} (Ref.~\cite{canopire}).}
\label{figure3}
\end{figure}

\begin{figure}
\centering
\includegraphics[width=0.5\linewidth]{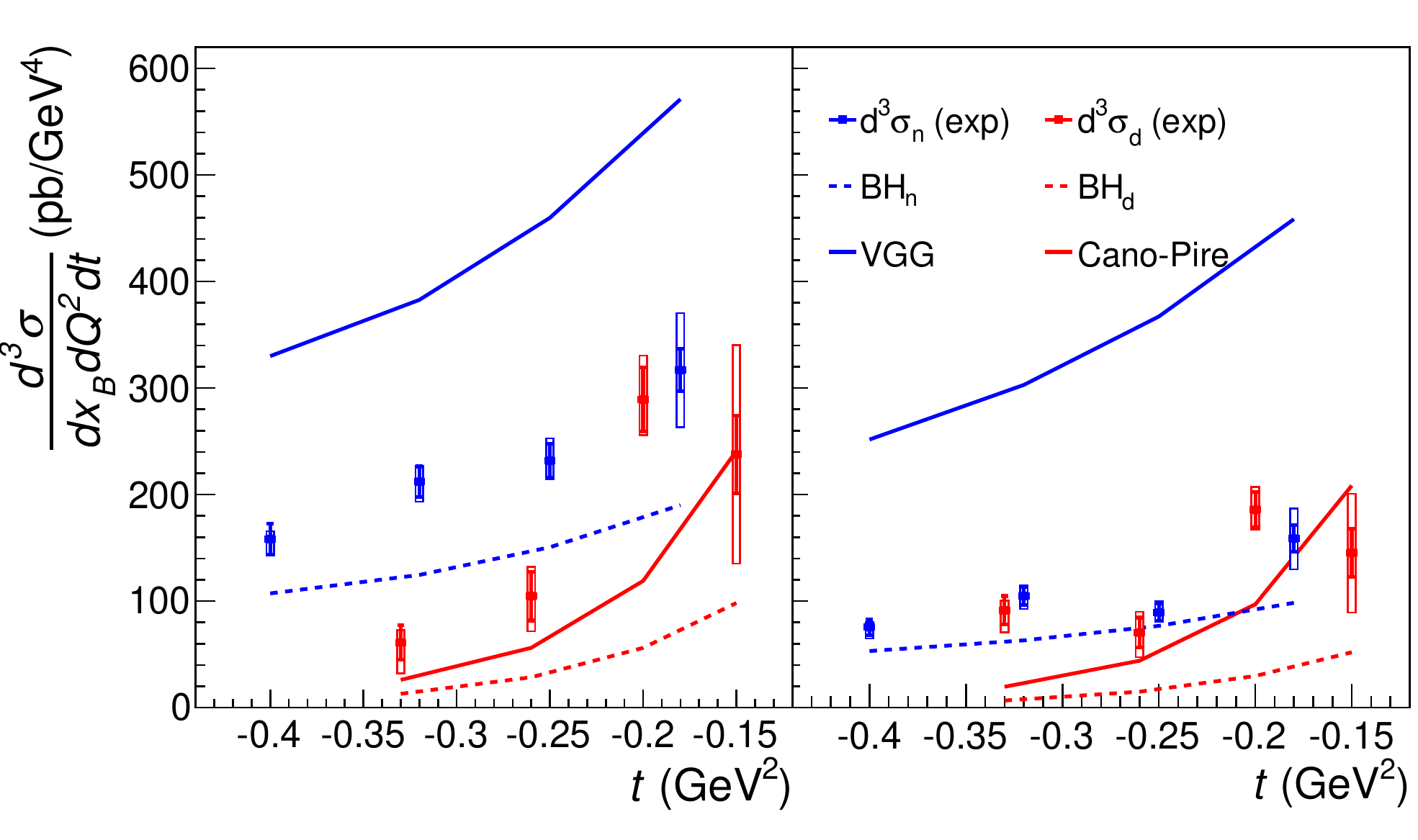}
\caption{Neutron and deuteron cross sections integrated over $\phi$. The blue (red) points correspond to the $e n \to e n \gamma$ ($e d \to e d \gamma$) experimental cross sections $d^3\sigma_{n (d)}/dQ^2dx_Bdt$ for beam energy $E=4.45$~GeV (left) and $E=5.55$~GeV (right). The error-bars show the s.d.~statistical uncertainty and the boxes around the points show the total s.d.~systematic uncertainty. Respective BH contributions are shown by the dashed lines, whereas the VGG model~\cite{Vanderhaeghen:1998uc, VGG} for the neutron and Cano-Pire model~\cite{canopire} for the deuteron are represented by the solid curves.}
\label{figure4}
\end{figure}

\begin{figure}
\centering
\includegraphics[width=0.5\linewidth]{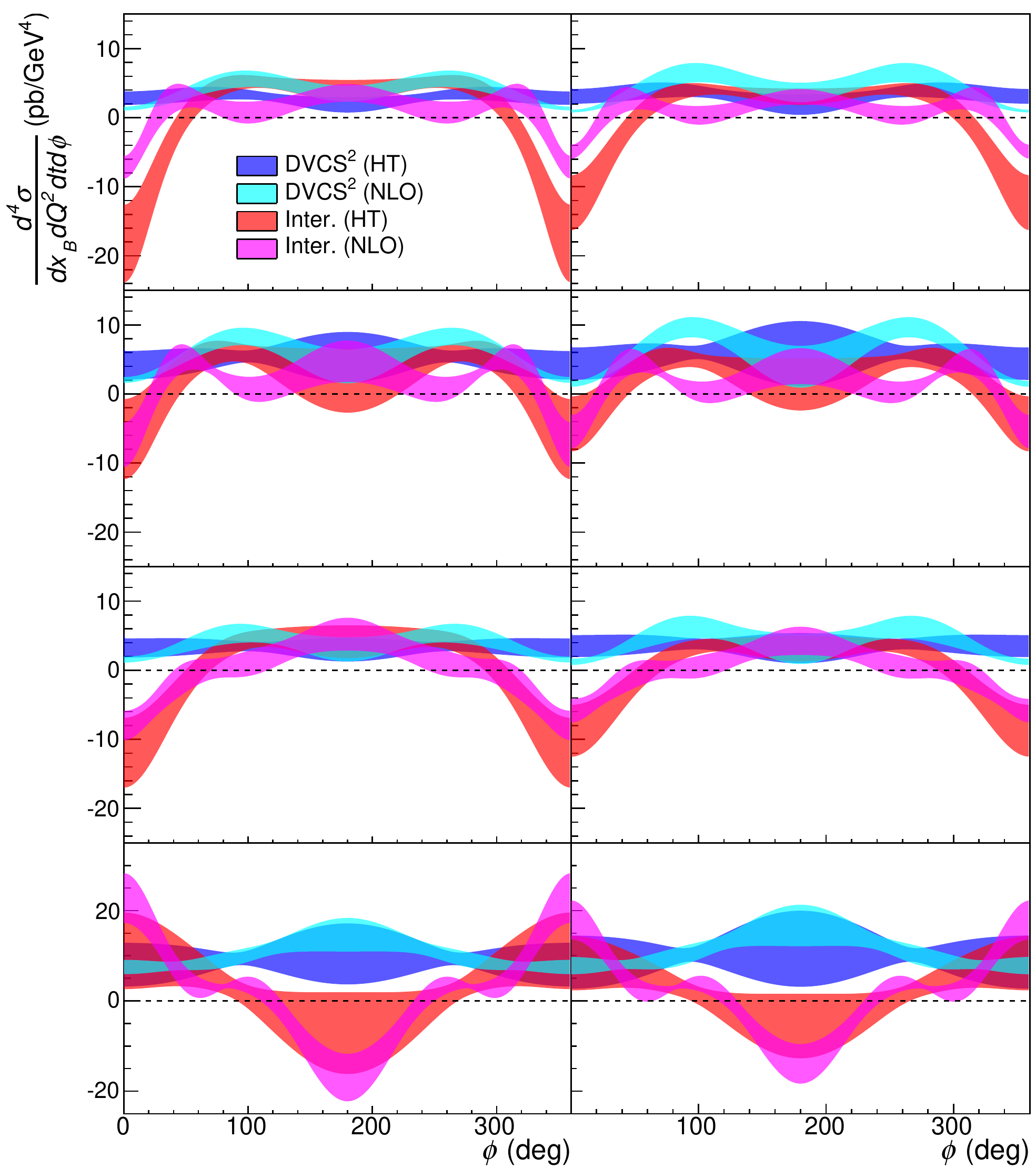}
\caption{Separation of the $|\mathcal{T}_{DVCS}^{n}|^2$ and $\mathcal{I}^{n}$ terms. The dark blue and light blue bands show the contribution to the $en\to en\gamma$ cross section of the $|\mathcal{T}_{DVCS}^{n}|^2$ term in the higher-twist (HT) and next-to-leading order (NLO) scenarios respectively. The red (magenta) band shows the contribution of $\mathcal{I}^{n}$ term in the HT (NLO) scenarios. The widths of the bands correspond to the s.d. statistical uncertainty of the fits. The results are for $E =4.45$ GeV (left) and $E=5.55$ GeV (right). From the top to the bottom, the squared momentum transfer corresponds to $-\left<t\right>=$0.40, 0.32, 0.25 and 0.18 GeV$^2$. The fits are performed within the cross-section formalism of Ref.~\cite{Braun:2014}.}
\label{figure5}
\end{figure}

\begin{figure}
\centering
\includegraphics[width=0.5\linewidth]{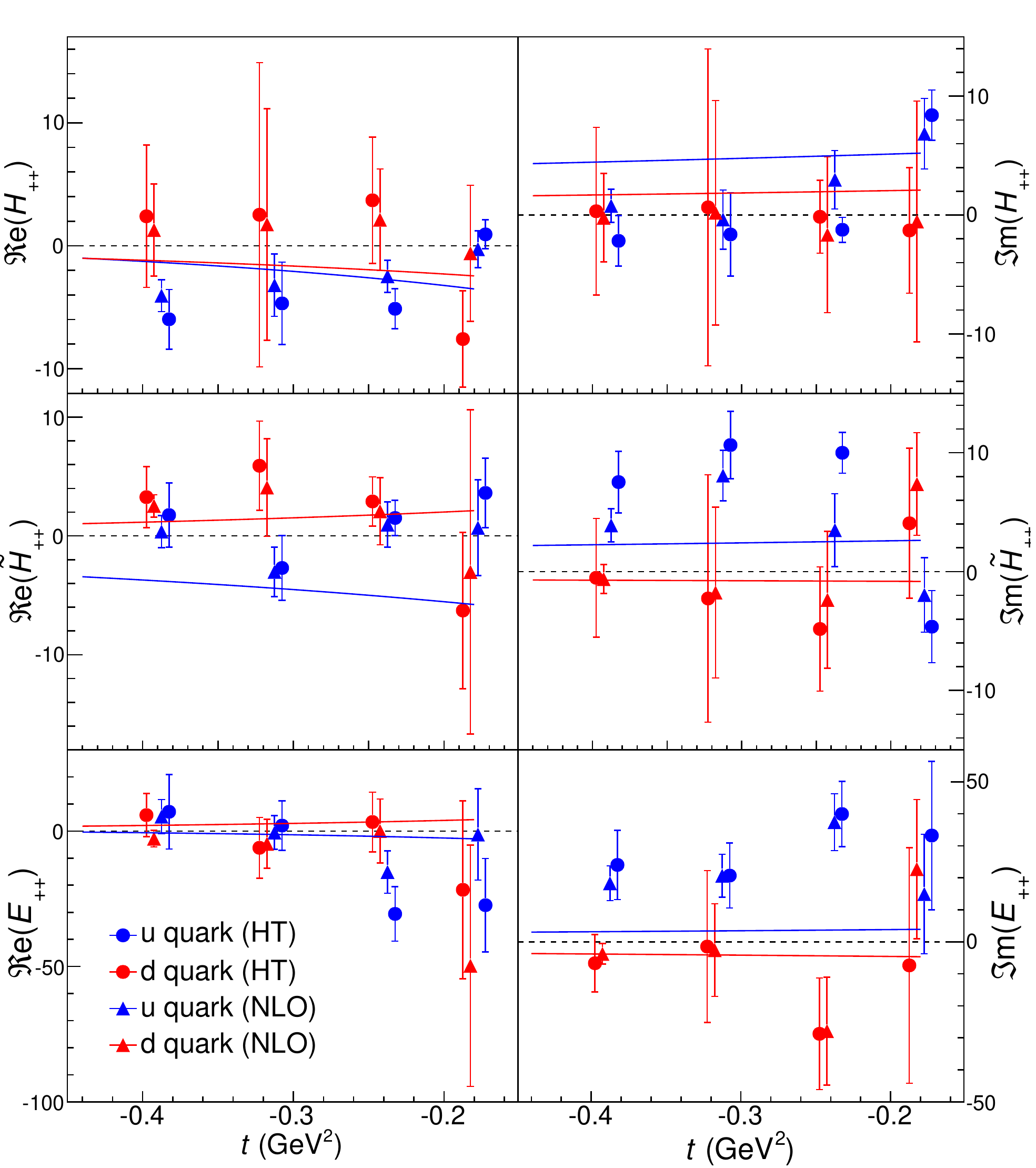}
\caption{Flavor separation of helicity-conserved CFFs. The points represent the real (left) and imaginary (right) parts of $u$ (blue) and $d$ (red) helicity-conserved CFFs $\mathcal{H}_{++}$, $\mathcal{\widetilde{H}}_{++}$ and $\mathcal{E}_{++}$ in both the higher-twist (HT) and next-to-leading order (NLO) scenarios. The error bars correspond to standard deviations and take into account the statistical and the systematic uncertainties of the fitted cross sections. Solid lines present the predictions of a reggeized diquark model of GPDs~\cite{Goldstein:2010gu, Goldstein:2013gra}.}
\label{figure6}
\end{figure}

The simulation used in the cross sections extraction is based on the GEANT4 toolkit. It takes into account the detector acceptance, the  
calculated pure Bethe-Heitler contributions from the neutron and deuteron, and the kinematic weights appearing in the cross section harmonics of the BH-DVCS interference terms and $|\mathcal{T}_{DVCS}|^2$ terms. 
Following the prescriptions in \cite{Vanderhaeghen:2000ws}, the simulation also includes the emission of hard photons (within the range of the missing mass squared spectrum of Fig.~\ref{figure2}) and we apply a correction factor to the extracted cross sections, to account for virtual photon  and soft-real photon emission.  This correction factor was 0.94 for
the specific neutron kinematics reported here.  We assign a 2\% systematic uncertainty to this factor, based on variations
of neutron \textit{vs.} proton,  varying  the neutron DVCS model in the radiative correction calculation, a very small
$\phi$-dependence, and differing procedures for exponentiating the soft photons. 
%of $\mathcal{I}^{n,d}$ and $|\mathcal{T}_{DVCS}^{n,d}|^2$. 
The calorimeter energy resolution in the simulation is smeared to fit the experimental one by reproducing the exclusive $M_X^2$ distribution of H$(e,e^{\prime}\gamma)X$ with $p(e,e^{\prime}\gamma)p$ simulated events. The obtained bin-by-bin smearing factors are then applied to the $d(e,e^{\prime}\gamma)d$ and the Fermi-smeared $n(e,e^{\prime}\gamma)n$ simulated data. This allows a proper computation of the experimental acceptance by applying identical selection criteria to experimental and simulated data, and to correct the final results for bin migration effects. This procedure introduces a bin-dependent systematic uncertainty which is added quadratically to a 3.1\% normalization uncertainty. The 3.1\% value originates from the uncertainties on the radiative corrections (2\%), the electron acceptance (1\%) and multi-track correction (0.5\%), the photon multi-cluster correction (0.5\%), the data acquisition deadtime and luminosity (2\%).

%Figures~\ref{apen:1} and \ref{apen:2} show the results of the fits presented above on the overall data set. Figure~\ref{apen:1} shows the fits on the proton data taken concurrently with the neutron data reported herein~\cite{Defurne:2017paw}, and Fig.~\ref{apen:2} shows data from a previous experiment at similar kinematics~\cite{Mazouz:2007aa,Defurne:2015kxq}. The values of $\chi^2/ndf$ of the fit are shown in Tab.~\ref{tab:chi2} for both the HT and NLO scenarios. The agreement is quite reasonable across the whole data set, including both helicity-dependent and helicity-independent cross sections off both the proton and neutron.

\bibliographystyle{naturemag}
\bibliography{Compton2018}

%\begin{table}[hb!]
%  \begin{tabular}{|l|r|}
%    \hline
%    Systematic uncertainty & Value\\
%    \hline
%    \hline
%    Electron acceptance & 1\%\\
%    Electron multi-track correction & 0.5\%\\
%    Photon multi-cluster correction & 0.5\%\\
%    Data acquisition deadtime and luminosity & 2\%\\
%    Radiative corrections & 2\%\\
%    \hline
%    \hline
%    Total:&3.1\%\\
%    \hline
%  \end{tabular}
%  \caption{Summary of normalization systematic uncertainties.}
%  \label{tab:sys}
%\end{table}

%\begin{table}[hb!]
%  \begin{tabular}{|l|c|c|c|c|}
%    \hline
%      $-\left<t\right>$ (GeV$^2$)& 0.18 & 0.25 & 0.32 & 0.40 \\
%     \hline
%   \hline
%     $\chi^2/ndf$ (HT) & $463/480$ & $521/480$ & $533/470$ & $399/444$ \\
%    \hline
%     $\chi^2/ndf$ (NLO) & $458/480$ & $530/480$ & $529/470$ & $407/444$ \\
%    \hline
%  \end{tabular}
%  \caption{Values of $\chi^2/ndf$ of the fit.}
%  \label{tab:chi2}
%\end{table}

\newpage

%\begin{figure}
%\centering 
%\includegraphics[width=1.\linewidth]{fig4.pdf}
%\caption{Subtraction of LH2 data from LD2 data. The histograms show the missing mass squared distributions of D$(e,e'\gamma)X$ (solid circles) and normalized Fermi-smeared H$(e,e'\gamma)X$  events (open circles) after accidental and $\pi^0$ background subtraction, integrated over $\phi$ ($E=4.45$ GeV and $\left<t\right>=-0.32$~GeV$^2$). Bars show standard deviation (s.d.) statistical uncertainties. The difference between the two distributions (squares) is scaled by a factor 10 for clarity. The blue and magenta bands (both scaled $\times 10$), show the simulated $n(e,e^{\prime}\gamma)n$  and $d(e,e^{\prime}\gamma)d$ yields, respectively, fit to the data by minimizing Eq.~(\ref{eq::fit}).  These bands include the s.d. statistical uncertainty of the fit. The total fit to the open squares distribution is shown with the solid (red) histogram. The vertical dashed lines delimit the exclusive region where the fit is performed.}
%\label{figure3b}
%\end{figure} 

%%%%%%%%%%%%%%%%%%%%%%%%
% EXTENDED DATA FIGURES
%%%%%%%%%%%%%%%%%%%%%%%%

\begin{figure*}
\centering
\includegraphics[width=\linewidth]{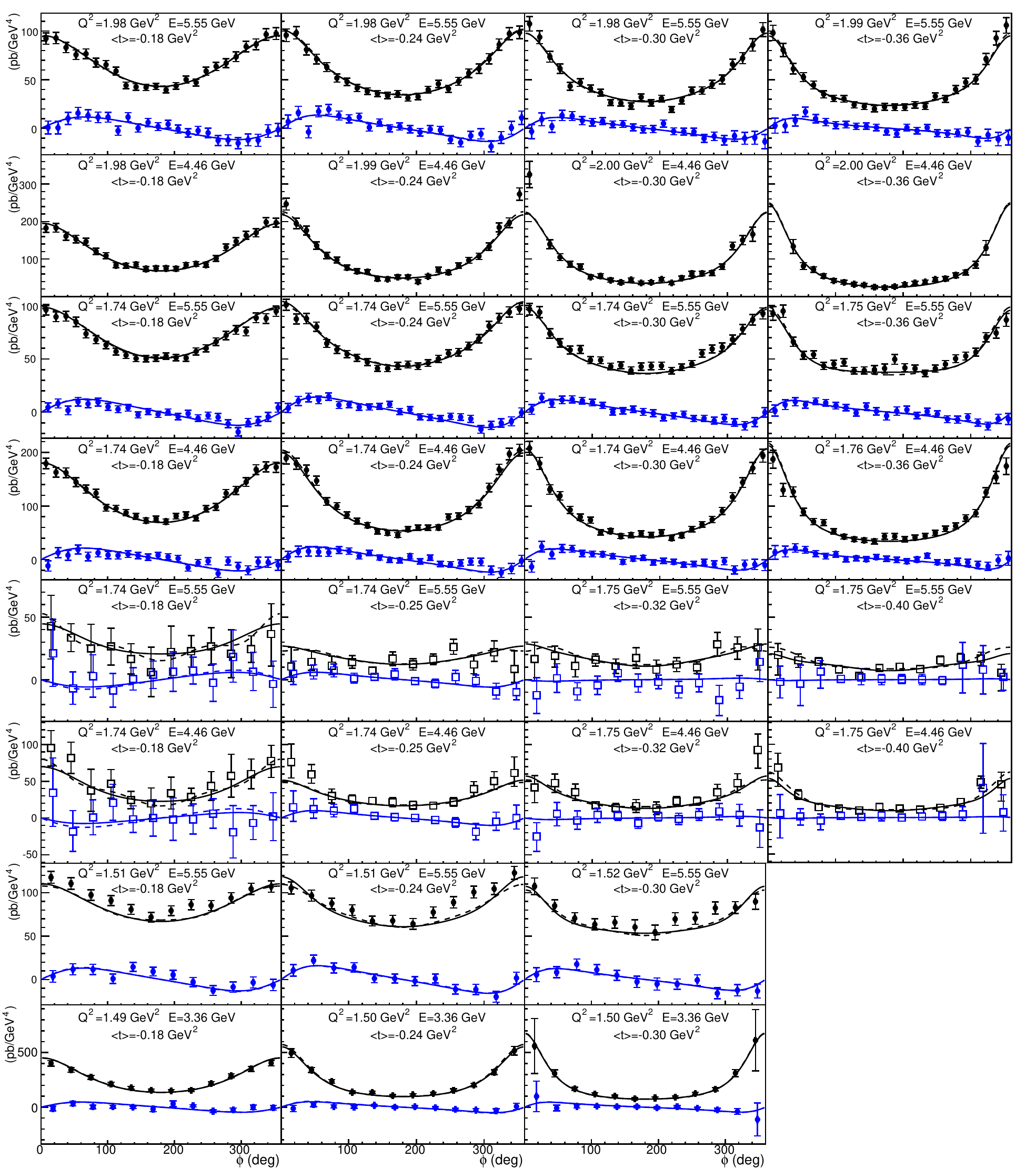}
\caption{Fit results on the 2010 data of E07-007 and E08-025 experiments. The plots show the helicity-independent (black) and helicity-dependent (blue) photon electro-production cross sections off proton (circles) and neutron (squares) from~\cite{Defurne:2017paw} and the data reported herein. The specific kinematics are indicated in each plot. Solid lines show the results of the HT fit described in this work, whereas the dashed lines (almost indistinguishable from the solid lines) show the results of the NLO fit.}
\label{apen:1}
\end{figure*}

\begin{figure*}
\centering
\includegraphics[width=\linewidth]{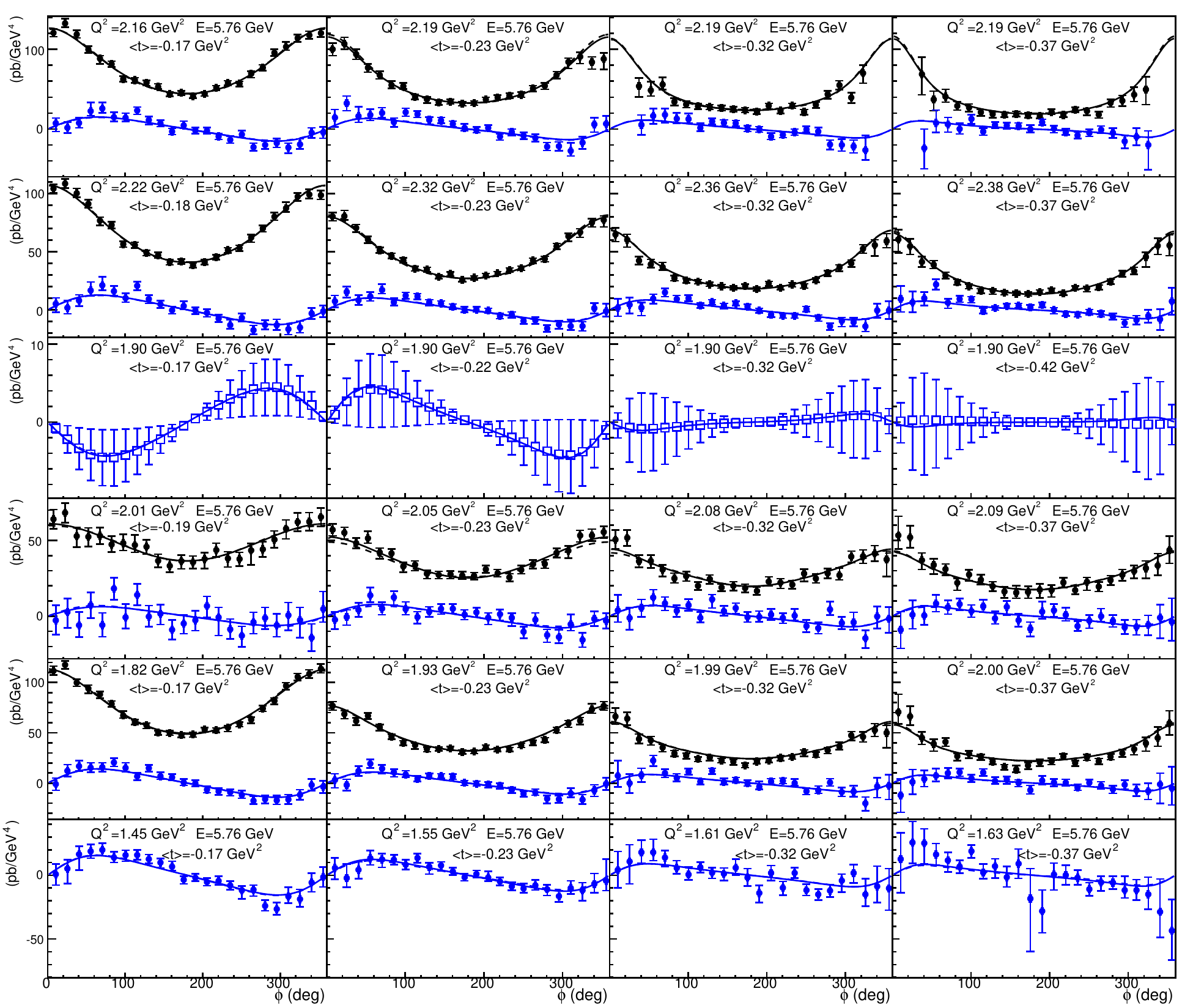}
\caption{Fit results on the 2004 data of E00-110 and E03-106 experiments. The plots show the helicity-independent (black) and helicity-dependent (blue) photon electro-production cross sections off proton (points) and neutron (squares) from~\cite{Mazouz:2007aa,Defurne:2015kxq}. The specific kinematics are indicated in each plot. Solid lines show the results of the HT fit described in this work, whereas the dashed lines (almost indistinguishable from the solid lines) show the results of the NLO fit. Neutron results in~\cite{Mazouz:2007aa} only contain the amplitude of the DVCS-BH interference term and its s.d. uncertainty. Data points in this figure for that experiment are placed along the calculated cross section, but without any spread around it.}
\label{apen:2}
\end{figure*}

%\bibstyle{apsref} 

\end{document}